\documentclass[a4paper,twoside]{article}
\newcommand{\affil}[1]{$^{\rm #1}$}

\newcommand{\omb}{\Omega_{\rm b}}
\newcommand{\omm}{\Omega_{\rm m}}
\newcommand{\oml}{\Omega_{\Lambda}}
\newcommand{\fb}{f_{\rm b}}
\newcommand{\ns}{n_{\rm s}}
\newcommand{\kms}{\,{\rm km}\, {\rm s}^{-1}}
\newcommand{\aeff}{A_{\rm eff}}
\newcommand{\tsys}{T_{\rm sys}}

\newcommand{\hmpc}{\,h^{-1}\,\rm Mpc}
\newcommand{\msol}{\,\rm M_{\odot}}

\newcommand{\mpch}{\,h\,\rm Mpc^{-1}}
\newcommand{\chisq}{\,\chi^{2}}

\usepackage[authoryear]{natbib}
\bibpunct{(}{)}{;}{a}{}{,}
\usepackage{graphicx}
\usepackage{rotating}
\usepackage{epsfig}
\usepackage{color}
\usepackage{float}
\usepackage{fixltx2e}
\usepackage{multirow}

\date{} 
\title{\large\bf\flushleft Cosmological surveys with the Australian Square Kilometre Array Pathfinder}
\author{\parbox{\textwidth}{\flushleft
\vspace{-0.5cm}
{\it Alan R. Duffy\affil{A}, Adam Moss\affil{B} and Lister Staveley-Smith\affil{A}}\\
\vspace{0.4cm}
{\small \affil{A}\,ICRAR, The University of Western Australia, M468, 35 Stirling Hwy, WA 6009, Australia}\\
{\small \affil{B}\,Department of Physics and Astronomy, University of British Columbia, 6224 Agricultural Road, Vancouver, BC, V6T 1Z1, Canada}}}
\begin{document}
\twocolumn[
\begin{changemargin}{.8cm}{.5cm}
\begin{minipage}{.9\textwidth}
\vspace{-1cm}
\maketitle
%
%
\small{\bf Abstract: This is a design study into the capabilities of the Australian Square Kilometre
 Array Pathfinder in performing a full-sky low redshift neutral hydrogen survey,
termed WALLABY, and the potential cosmological constraints one can attain
from measurement of the galaxy power spectrum.  We find that the full sky survey will likely 
attain $6 \times 10^5$ redshifts which, when combined with expected {\em Planck} CMB data, will constrain
the Dark Energy equation of state to 20\%, representing a coming of age for radio observations in creating
cosmological constraints.}

\medskip
\medskip
\end{minipage}
\end{changemargin}
]
\small

\section{Introduction}
\label{Introduction}

With the advent of large cosmological volume galaxy surveys, comprised of well measured 
positional information from homogenous datasets, the measurement of the galaxy matter power 
spectrum has become almost routine.
The use of such power spectra in the determination of the cosmological model has
been based almost exclusively, however, on optical techniques, e.g. 
the 2dF Galaxy Redshift Survey (2dFGRS\footnote{2dF homepage: www.aao.gov.au/2dF})
and the Sloan Digital Sky Survey (SDSS\footnote{SDSS homepage: www.sdss.org}). 
The overall shape of the power spectrum is sensitive to the cosmological parameter constraints 
$\Gamma = \omm\,h$ and $\fb=\omb / \omm$, where $\omm$ and $\omb$ are 
the total matter and baryon densities defined relative to critical, and 

\noindent $h=H_0/(100 \, \kms\,{\rm Mpc}^{-1})$, as well as the spectral index of the density fluctuations, $\ns$, 
and neutrino densities~\citep[e.g.][]{Percival, Tegmark2004a, Tegmark2004b, Cole:05, AR:07}.
Additional information from the power spectrum can be gleaned by measuring the physical scale of the so-called 
Baryonic Acoustic Oscillations~\citep[e.g.][]{Blake, Percival10} either through using the scale as a `standard ruler'
or by combining with measurement of the Cosmic Microwave Background to break degeneracies. These measurements
allow the determination of the nature of the Dark Energy,  through constraining the equation of state parameter, $w$, 
which for the cosmological constant is $-1$. With enough sufficiently distant galaxies the variation of the 
equation of state parameter with redshift can be constrained by comparing this acoustic scale in different epochs,
as considered in, e.g.~\citet{AR:10}.

Recent advances of the speed at which radio telescopes can survey the sky to a given flux limit
point to the possibility of radio joining optical surveys to measure the matter power spectrum.
The distribution of these sources along the line of sight is accurately determined by using the redshifted emission line 
at $\approx 21\,\rm cm$ of the hyperfine splitting transition in neutral hydrogen (HI). 
Previously surveys have been limited to $\sim 10^3$ galaxies~\citep[e.g.][]{hipass,hijass} 
whilst the very latest HI catalogue from the Arecibo legacy survey, ALFALFA, is expected to find 
$\sim 10^4$ objects~\citep{ALFALFA}. In the near future the Chinese-built Five-hundred Aperture Spherical 
Telescope~\citep{KARST} could detect as many as $\sim 10^{6}$ in the current design~\citep{Duffy:08a}. Ultimately 
however, the future for radio galaxy surveys is the Square Kilometre Array, 
SKA\footnote{SKA homepage: www.skatelescope.org}, which may detect $\sim 10^9$ 
galaxies~\citep{AR}. The initial step towards the SKA facility is a precursor known as the Australian SKA Pathfinder or 
ASKAP\footnote{ASKAP homepage: www.askap.org}.
The pathfinder consists of a much reduced number of telescopes, but still operating with a large Field of View (FoV) of the sky,
which therefore enables the revolutionary upgrade in survey speed.

The low-redshift precursor surveys of the SKA will accurately measure the properties of galaxies at low redshift and how these 
properties change as a function of environment, for example what is the dependence of the HI mass function on local galaxy density? 
The deeper precursor surveys will measure evolutionary effects. 
In this work we will demonstrate that simple estimates of the number and distribution of HI detected galaxies will 
enable ASKAP to be the first radio telescope to derive cosmological parameter constraints, able to constrain the Dark Energy equation
of state to 20\%. This is a similar capability to previous optically based measurements such as with
2dF~\citep{Cole:05} but significantly more collecting area will be needed to rival current optical surveys
such the 6dF~\citep{Beutler:11}, SDSS II luminous red galaxy survey~\citep[e.g.][]{Thomas:11} and WiggleZ~\citep{Blake:10}
surveys much less forthcoming optical surveys such as the SDSS III Baryon Oscillation Spectroscopic Survey 
(BOSS\footnote{BOSS homepage: www.sdss3.org/surveys/boss.php}).
However we note that future radio surveys with, for example, the SKA have the capability to exceed the volume and numbers of galaxies 
{\it spectroscopically} detected compared with optical surveys.  

While limiting our study to the use of the power spectrum in constraining 
cosmology~\citep[e.g.][]{Blake} we note that the spectroscopic nature of radio surveys enable other cosmological 
probes. One such use is in measuring redshift-space distortions, a statistical measurement of infalling galaxies near large scale structure, 
which can probe the growth of structure on cosmic scales and hence providing a strong test of the validity of 
General Relativity as well as Dark Energy models~\citep[e.g.][]{Song:09,Blake:10} over $\sim$Mpc scales.
With a relatively high density of galaxies probing a given volume WALLABY will likely equal, if not exceed, constraints on this measurement 
by an optical survey such as 6dF (Beutler et al., {\it in prep}).
Instead of using the redshifts one can calculate the distance to the galaxies themselves, using relations such as the 
Tully-Fisher~\citep{Tully:77} or Fundamental Plane~\citep{Faber:87,Djorgovski:87}, 
to estimate the Hubble flow which can then be removed to leave the peculiar velocities 
of each galaxy. These velocities form a velocity field, or Bulk Flow, that reflects the matter distribution, allowing constraints of the amount of 
mass and the fluctuation of overdensities in the Universe~\citep[e.g.][]{Burkey:04,Abate:08,Watkins:09}.

We detail the techniques and assumptions considered in our calculation of galaxy detections in 
Section~\ref{sec:method}, in particular the effect of telescope resolution and galaxy inclinations in limiting galaxy counts (Section~\ref{sec:resolve}). Utilising these assumptions we calculate the expected number of galaxies that the all-sky ASKAP (WALLABY) survey
could be expected to find in Section~\ref{sec:galsurvey}. By constraining the matter 
power spectrum we then estimate the suitability of the WALLABY survey as a cosmological probe in 
Section~\ref{sec:cosmo}.

\section{Method}
\label{sec:method}

We have utilised a, significantly, updated methodology to~\cite{Duffy:08a} which analysed
the potential galaxy surveying power of the Five hundred metre Aperture 
Spherical Telescope (FAST). Therefore the reader may wish to consult that article for a more
in-depth discussion on the following issues, including a consideration of evolution in the HI mass
function (WALLABY is a shallow survey and hence likely to be unaffected by evolution). However
there is one significant difference between FAST and ASKAP, namely that the former is a single 
dish and the latter an interferometer. A difference that potentially has significant effects in terms of 
resolving out extended structure. As we shall see the loss of signal from this effect is an issue for 
objects at all redshifts not just those closest to the observer (and hence with the largest angular extent on the sky).
In other words, with the large baselines, $2\,\rm km$, available to WALLABY, 
most detections are resolved and hence one must consider this issue. The positive counterpoint to this
high resolution is that the galaxies rarely overlap within the beam of the telescope and the survey is effectively
never confusion limited, as discussed in Section~\ref{sec:confusion}. 

For an interferometer observing a resolved structure one can attempt to recover some (but not all) of the missing flux by smoothing the 
data, i.e. spatially integrating. However, some flux will be lost as spatial smoothing is akin to removing the longer antenna baseline pairs 
and therefore results in a loss of sensitivity. 
Exact characterisation of the effect depends in detail on the distribution of neutral hydrogen in each 
galaxy and the properties of the `source finder'. 
This is a complex issue and a detailed discussion of angular size distributions of detected galaxies is therefore 
deferred to a later paper (Duffy et al, {in prep}). 
However, a simple approximation is that, if there are $n$ independent pixels 
(beams) that make up a galaxy (where we use the example $n>>1$), the S/N ratio of those pixels, when combined, will be 
improved by $\sqrt{n}$ if the object is approximated by a top hat column density and velocity profile.
We consider this effect in more detail in Section~\ref{sec:resolve} and find that there is a non-negligible reduction in the galaxy
counts of order 20\% if one considers a reduction in S/N due to this effect.

\subsection{Estimating the HI signal}
As detailed in~\citet{Duffy:08a} and references therein, the expected thermal noise for a dual 
polarisation, $n_{\rm pol} =2$, single beam is given by
\begin{eqnarray}\label{eq:flux noise} 
\sigma_{\rm noise}=\sqrt{2}\frac{k\tsys}{\aeff}\frac{1}{\sqrt{n_{\rm pol} \Delta \nu \,t}}, \end{eqnarray}
for an observing time of $t$ and a frequency bandwidth $\Delta \nu$, where 
$k=1380\,{\rm Jy}\,{\rm m}^2\,{\rm K}^{-1}$ is the Boltzmann constant and $T_{\rm sys}$ is the system temperature (assumed to be 
$50\,\rm K$). 
The effective area, $A_{\rm eff}$, calculation has been modified from the previous single dish calculation to better
reflect the interferometer nature of ASKAP. The individual effective area of an 
ASKAP dish is the geometric area of a $12\,\rm m$ diameter dish, $a_{\rm eff}$, reduced by the aperture 
efficiency, expected to be $\alpha_{\rm eff} \approx 0.8$~\citep{Johnston:08}. 
Due to computational limitations in correlating signals from all 36 dishes in ASKAP, WALLABY will likely use the inner
core of 30 dishes, $N_{\rm dish} = 30$, which can be combined in 
$N_{\rm perm} = N_{\rm dish} (N_{\rm dish}-1) / 2$ permutations. The resolution of the inner core is 
limited to $30''$ at 21cm wavelength using the central $2\,\rm km$ baselines of ASKAP.
For each pairwise correlation we assume a $\sqrt{2}$ boost to the signal-to-noise by averaging the real 
and imaginary signal from a complex correlator~\citep{Thompson:99}.
This leads to an overall effective area for ASKAP of 
\begin{eqnarray}\label{eq:flux noise beam} 
\sigma_{\rm noise}=\sqrt{2}\frac{k\tsys}{ \alpha_{\rm eff} a_{\rm eff}}\frac{1}{\sqrt{n_{\rm pol} N_{\rm dish} (N_{\rm dish}-1) \Delta \nu \,t}}\,, \end{eqnarray}
where we have averaged over the complex and real signals.

Typically, the beam area increases like $\lambda^{2}\propto (1+z)^2$ which, if one uniformly tiles 
the $z=0$ sky, has the positive result that slices at higher redshift receive extra exposure due 
to the fact that observations will overlap. This reduces the flux limit relevant to a particular 
redshift slice by a factor $(1+z)^{-1}$, as discussed by~\citet{AR}. This is not the case for ASKAP 
however, as the number of on sky beams varies as a function of frequency to ensure that there is
an approximately fixed covering area as a function of redshift.
Hence, the flux limit for an observation, $S_{\rm lim}$, for a specific signal-to-noise ratio $(S/N)$ is 
given by
\begin{eqnarray}\label{eq:fluxlimit} S_{\rm lim}= (S/N) \sigma_{\rm noise}\,.\end{eqnarray}
We relate this flux to the HI mass, $M_{\rm HI}$, of a galaxy at redshift $z$ 
in terms of the observed flux, $S$, and line width, $\Delta V_{\rm o}$, 
by~\citet{Roberts}
\begin{eqnarray}
\label{eq:mass limit} 
\frac{M_{\rm HI}}{M_{\odot}} = \frac{2.35 \times 10^{5}}{1+z} \left( \frac{d_{\rm L}(z)}{\rm Mpc} \right)^{2} \left(\frac{S}{\rm Jy} \right) \left(\frac{\Delta V_{\rm o}}{\kms} \right), 
\end{eqnarray}
where $d_{\rm L}(z)$ is the luminosity distance to the galaxy, necessitating the $(1+z)^{-1}$
correction for an FRW universe. In a significant departure from the methodology of~\citet{Duffy:08a} 
we make use of the measured number density of objects as a function of velocity widths and HI masses
directly from HIPASS, presented in~\citet{Zwaan:10}. With this method we automatically include
the effects of angle of inclinations of galaxies as well as the complex velocity-structure of the system.

\begin{figure}[!t]
  \begin{center}
    \epsfysize=2in
    \epsfxsize=4in
    \epsfig{figure=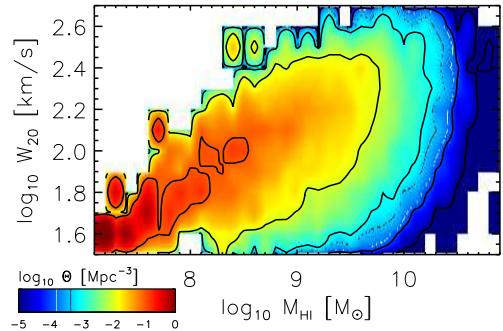, scale=0.4}
    \caption{We have created a matrix of the detections from the HIPASS survey as a function of velocity widths 
    		and inferred HI mass. This matrix currently represents the latest understanding in the distribution of 
		HI detected galaxies, with angle of inclination effects as well as rotation - mass relations represented.}
    \label{fig:HIPASS_W20_M}
  \end{center}
\end{figure}

In Fig.~\ref{fig:HIPASS_W20_M} we show the full matrix utilised noting that the histogram widths are 
0.01 dex whereas the colour scheme is the standard number density in decades of mass and velocity.
We emphasise that this represents the very latest information pertaining to the frequency of HI systems
as a function of mass and velocity widths and, due to the limited redshift surveyed by WALLABY, is an
ideal basis for estimating galaxy number counts.

To explicitly incorporate the matrix information we must recast Eq.~\ref{eq:mass limit}, first dividing by $\Delta V_{\rm o}$ 
(the ratio is termed a peak flux $S^{\rm peak}\equiv M_{\rm HI} / \Delta V_{\rm o}$, implicitly assuming the galaxy spectral profile
is a top-hat). This means that for each grid cell in
Fig.~\ref{fig:HIPASS_W20_M} we can also divide the HI mass by the velocity width ($W_{20}$ is assumed to be the full velocity
width of the system) and read off the number density of galaxies
in those cells which lie above this peak flux. To determine what the minimum observable flux is (as a function of redshift) 
we substitute the flux limit of the survey (Eq.~\ref{eq:fluxlimit}) for $S$ in Eq.~\ref{eq:mass limit} allowing us to create the following
inequality to determine when a population of HI sources will be detected by ASKAP
\begin{eqnarray}
\label{eq:peak limit} 
S^{\rm peak} & \equiv & \frac{M_{\rm HI}}{M_{\odot}} \left(\frac{\Delta V_{\rm o}}{\kms} \right)^{-1} \sqrt{N_{\rm ch}} \nonumber \\
& \ge & \frac{2.35 \times 10^{5}}{1+z} \left( \frac{d_{\rm L}(z)}{\rm Mpc} \right)^{2} \left(\frac{ (S/N) \sigma_{\rm noise}}{\rm Jy} \right) \,,
\end{eqnarray}
where $N_{\rm ch}$ is the number of channels that the galaxy will be distributed across, calculated as rounding to the nearest integer 
number of channel $\Delta V_{\rm o} / dV$ (where dV in ASKAP is $4\kms$) in a calculation similar to Eq. 8 of~\citet{AR}.

One can estimate the number of galaxies detected in the survey by adding the densities of all populations that fulfil the inequality in 
Eq.~\ref{eq:peak limit} with a peak flux $S^{\rm peak}_{\rm lim}(z)$ by computing
\begin{eqnarray}\label{eq:zeroth} N\left( M > M_{\rm lim},z\right) = \Delta\Omega \Delta z \frac{dV}{dzd\Omega} \int_{S^{\rm peak}_{\rm lim} \left( z \right)}^{\infty}\,\frac{dN}{dVdS^{\rm peak}}\,dS, \end{eqnarray}
where the  sky area covered is  $\Delta\Omega$ and the size of the redshift bin is $\Delta z$, $dV/dzd\Omega$ is the comoving volume element for the 
FRW universe and $dN/dVdS^{\rm peak} \equiv dN/dV(dM_{\rm HI}/dW_{20})$ is the comoving number density of galaxies per unit peak flux, 
taken from the matrix of Fig.~\ref{fig:HIPASS_W20_M}.
No evolution in the HI mass-velocity width space has been assumed over the limited redshift surveyed in WALLABY, nor is there
any apparent evolution in the integrated cosmic HI density out to $z\approx 0.8$~\citep{Chang:10} or indeed
$z \approx 2$~\citep{Prochaska:10}. We calculate the average redshift of galaxies in the survey from 
$N(S^{\rm peak}>S^{\rm peak}_{\rm lim},z)$ by integrating appropriately over $z$, that is, 
\begin{eqnarray}
\langle z\rangle = {\int_0^{\infty}z\,N(S^{\rm peak}>S^{\rm peak}_{\rm lim},z)\,dz
\over \int_0^{\infty}N(S^{\rm peak}>S^{\rm peak}_{\rm lim},z) \,dz}\,.
\end{eqnarray}

\begin{table}[!h]
\begin{center}
\begin{tabular}{cc} \hline
Parameter & ${\rm WALLABY}$ \\ \hline
$\aeff$ (${\rm m^{2}}$) & 2668 \\
$\tsys$ ($K$)& 50 \\
Maximum Baseline (km) & 2 \\
Angular Resolution (z=0) & 30''  \\
Sky Coverage (${\rm deg^{2}}$) & 30000 \\
Total Survey Time (hrs) & 9600 \\
Redshift range & 0 - 0.26  \\ 
Total number of galaxies & 673321 (848566)  \\
Mean redshift of sample & 0.0492 (0.0557)  \\
$V_{\rm eff}$ at $k=0.065\hmpc$ & $6.9\times 10^{7} \, \rm Mpc^{3}$ \\ 
$\sigma_{P}/P$ at $k=0.065\hmpc$ & $9\%$ \\ 
\hline
\end{tabular}
\medskip\\
\label{tab:survey_values} 
\caption{We summarise here the survey specific values of WALLABY~\citep{WALLABY}
in addition to the strawman values of ASKAP~\citep{Johnston:08}. We consider the
reduced baseline model for WALLABY which utilises the inner 30 dishes across 
a maximum $2\, \rm km$ baseline rather than the full 36 dish, $6\,\rm km$ extent of ASKAP.
We also have two numbers for the predicted galaxy counts, and their mean redshift, reflecting
the effects of including the reduction of signal-to-noise by spatially resolved galaxies, as
demonstrated in Fig.~\ref{fig:ASKAP_mfn_dndz}. The brackets ignore this effect and hence have a larger 
galaxy count. We consider the conservative estimate for ASKAP when noting the effective volume, 
as given in Equation~\ref{eq:veff_const}, for the typical $k$-mode of interest (roughly the position of the first
Baryonic Acoustic Oscillation peak). We also give the estimated percentage 
measurement error of the power spectrum at this scale. 
The maximum redshift we can probe the first peak out to, with $nP=3$, is $z=0.116$ with a number density of sources 
estimated at $n=1.07\times 10^{-4} \,\rm Mpc^{-3}$. At the second peak $k=0.125 \hmpc$
we find the effective volume to be $3.6 \times 10^{7} \, \rm Mpc^{3}$ and $\sigma_{P}/P = 7\%$.}
\end{center}
\end{table}

\subsection{Confusion of galaxies}\label{sec:confusion}

A limiting factor in galaxy surveys is the issue of confusion, whereby detections in HI are unable to be unambiguously assigned to a 
single galaxy. Typically HI surveys have previously had far greater discrimination between objects along the line of sight than in the plane of the sky. ASKAP will differ in this regard by enabling wide-field surveys of the sky with at least 30'' resolution (ASKAP baselines of 
$2\, \rm km$) together with highly competitive $4 \kms$ velocity resolution. It is therefore unlikely that confusion will play a significant role 
in limiting the number of galaxy detections in the WALLABY survey, an expectation we verify by making use of a simple analytic estimate 
of the expected level of confusion (as given in~\citealt{Staveley-Smith:08}). 

To estimate the confusion rates one must have a measure of how many galaxies are in the survey volume, which we estimate from the 
HI mass function as measured by~\citet{hipass},
\begin{equation}\label{eqn:schechter}
\phi(M_{\rm HI}) d M_{\rm HI} = \Theta^{\star} \left( \frac{M_{\rm HI}}{M^{\star}} \right)^{\alpha} \exp \left( \frac{M_{\rm HI}}{M^{\star}} \right) d \left(\frac{M_{\rm HI}}{M^{\star}} \right)\,,
\end{equation} 
where $\alpha = -1.37$, $\Theta^{\star} = 1.42 \times 10^{-2} (\hmpc)^{-3}$ and $M^{\star} = 10^{9.8} \msol$.
The number density of galaxies, $n_{0}$, with a mass larger than $M_{\rm lim}$ is given by the integral of this Schechter function
resulting in the well known $\Gamma$ function
\begin{equation}\label{eqn:intschechter}
n_{0}(M > M_{\rm lim}) = \Theta^{\star} \Gamma \left(1+\alpha, \frac{ M_{\rm lim} }{ M^{\star} } \right)\,.
\end{equation}
The total HI mass contained in systems above $M_{\rm lim}$ is given by the mass weighted integral of the Schechter function
\begin{equation}\label{eqn:mintschechter}
\int^{\infty}_{M_{\rm lim}} M_{\rm HI} \phi(M_{\rm HI}) d M_{\rm HI} = M^{\star} \Theta^{\star} \Gamma \left(2+\alpha, \frac{ M_{\rm lim} }{ M^{\star} } \right)\,.
\end{equation}
We are interested in understanding the confusion rates with sources that would significantly contribute to the mass of the main detection. 
Using Eqn.~\ref{eqn:mintschechter} we can evaluate the HI mass fraction contained in sources above a limiting mass; only $21\%$ of the HI 
in the Universe is contained in systems of mass greater than $M^{\star}$, while $75\%$ of the mass is contained within systems more massive
than $0.1 M^{\star}$. To a good approximation therefore we can calculate confusion rates from sources of this mass and greater. The typical
density of sources above $0.1 M^{\star}$ is found using~\ref{eqn:intschechter}, which for the HI mass function of~\citet{hipass} is 
$0.017 \, \rm Mpc^{-3}$.

We then calculate the typical distribution of the galaxies on the sky, as given by the galaxy-galaxy distribution which relates the typical
density of sources $n_0$ as a function of comoving distance $r$ from a given galaxy, approximated in the non-linear regime by
\begin{equation}
\rho(r) = n_{0} \left( 1+\left(\frac{r}{r_{0}} \right)^{\gamma} \right)\,,
\end{equation}
where $r_{0}$ is the correlation length. The average number of objects in a cylinder of comoving line-of-sight depth, $\beta$,
and transverse comoving radius, $\kappa$, is
\begin{equation}
n(\beta,\kappa) = \int \int 2\pi \kappa \rho(r) d\kappa d\beta\,.
\end{equation}
For $r^2 = \kappa^2 + \beta^2$ the solution is 
\begin{equation}
n(\beta,\kappa) = \pi \beta \kappa^2 n_{0} \left(1-\frac{2}{\gamma - 2}\left(\frac{r_0}{\kappa} \right)^{\gamma} \, _{2}F_{1}\left(\frac{1}{2}, \frac{\gamma}{2} -1;\frac{3}{2}; -\frac{\beta^2}{\kappa^2} \right)   \right)\,,
\end{equation}
where $_2 F_1$ is a hypergeometric function. Assuming, reasonably, that the distribution of galaxies does not evolve over the redshift
range probed by WALLABY we can make use of the HIPASS galaxy-galaxy correlation function measurements 
of~\citet{Meyer:07} ($\gamma = -1.5$ and $r_0 = 4.7 \,\rm Mpc$).
The cylinder diameter, $2\kappa$, is set by the ASKAP beam. For the natural Gaussian antenna distributions described 
in~\citet{Staveley-Smith:06} and modelled in~\citet{Gupta:08} the Full-Width Half Maximum beam extent for WALLABY is 
\begin{equation}\label{eq:beamdiam}
\Omega_{\rm FWHM} = 1.4 \lambda / 2000\,\rm m\,,
\end{equation}
where we have conservatively assumed that the maximum ASKAP baseline is $2\, \rm km$.
The cylinder depth is set by the accuracy of the available redshifts for the confusing population, $\Delta z$.

Although the $4\kms$ velocity width of ASKAP will ensure that the spectroscopic redshifts of WALLABY will be measured
to approximately $10^{-5}$ the best spectroscopic redshift estimate in practice will be limited by the typical Doppler width of the galaxy.
If we assume, conservatively, that two $M^{\star}$ galaxies of typical velocity width $300\kms$ 
are {\it just} overlapping then $\Delta z = 600 \kms /c = 0.002$. 
Even with this conservative calculation the chances that there will be more than one galaxy in the beam volume, i.e. the confusion rate,
is at a negligible sub-percent level for the mean redshift of the survey $z\approx 0.05$  (determined in Section~\ref{sec:galsurvey}) 
where the majority of detections lie.
However, as one surveys deeper in the Universe the confusion rate will steadily increase as the telescope beam encompasses greater regions of
space, further compounded by the lengthening observed wavelength, yet 
even at the survey edge of WALLABY, $z=0.26$ we find that less than 5\% of galaxies will be confused.

Using this formalism we can also calculate the typical success rates in assigning an optical counterpart to the HI detected galaxies, 
assuming that such an optical catalogue contains all these HI galaxies we can then utilise the same number density as before. 
As a worst case scenario we further assume that only optical photometric redshifts are available, 
with a `typical' redshift error $\Delta z\approx 0.05$~\citep{Hildebrandt:08}. Thus the chances that we 
can unambiguously assign an optical counterpart (i.e. there is only one galaxy in the volume) occurs in more than 97\% of cases at the average
redshift of the WALLABY survey $z \approx 0.05$ (determined in Section~\ref{sec:galsurvey}). At the survey edge the success rate is 80\% 
which is more than acceptable for most science cases. However, this value is a conservative case as we could blindly assign a counterpart
from the candidates, i.e. if there were two galaxies in the beam volume then we would be right 50\% of the time. This could be further improved
by using prior knowledge and choosing the largest stellar counterpart, for example.

In conclusion, provided ASKAP baselines of $2\, \rm km$ are available (i.e. the most conservative case) 
the overall galaxy number counts will be largely unaffected by confusion and this effect is henceforth ignored in the following discussion.
To enable unambiguous optical counterparts detections for the majority of ASKAP HI detections we find that photometric errors of order 
$\Delta z \approx 0.05$ are sufficient unless surveys probe deeper in redshift (by $z\approx 0.4$ the success rate is less than 50\%) 
or precise redshifts are needed, for example if attempting spectral stacking experiments, in which case a spectroscopic follow up in the
optical is demanded.

\subsection{Resolving rotating galaxies}\label{sec:resolve}
An important consideration for interferometers is the issue of resolving out galaxies that are larger
in extent than the beamsize. For ASKAP, with a $2 \, \rm km$ baseline this will certainly be an
issue for extended sources. 
Furthermore with $4\kms$ velocity resolution ASKAP will also probe the velocity structure of the 
galaxies themselves hence we try to estimate the size, inclination and typical rotation velocities
of the galaxies that are present in the HI mass-velocity matrix.

We estimate the angle of inclination, $\theta$, of the galaxy by relating the measured velocity width $\Delta V_{\rm o}$ from the HI-mass-velocity matrix, 
to the intrinsic linewidth width, $\Delta V_{\rm e}$. The difference between the true velocity and observed velocities is proportional to $\sin(\theta)$. We can
use the HI mass from the HI mass-velocity matrix to determine the intrinsic linewidth of a galaxy, corrected for broadening, using the empirical relation found
by~\citep{BriggsRao:1993, hijass} to be
\begin{eqnarray}
\label{eq:velocity-mass relation}
\frac{\Delta V_{\rm e}}{420\,{\rm km}\,s^{-1}}=\left( \frac{M_{\rm HI}}{10^{10}M_{\odot}} \right)^{0.3}\,, \end{eqnarray} 
although we note that this relation shows a large dispersion, especially for dwarf galaxies. 
The linewidth of a galaxy, $\Delta V_\theta$, which subtends an angle $\theta$ between 
its spin axis and the line-of-sight can be computed using the Tully-Fouque rotation 
scheme~\citep{TFq}
\begin{eqnarray}\label{eq:TFq}
\lefteqn{({\Delta V_{\rm e} \sin(\theta)})^{2} = (\Delta V_{\theta})^{2} + (\Delta V_{\rm t})^{2}  - }\nonumber \\
& & 2{\Delta V_{\theta}}{\Delta V_{\rm t}}\left( 1- e^{- \left(\frac{{\Delta V_{\theta}}}{\Delta V_{\rm c}}\right)^{2}} \right) - 2(\Delta V_{\rm t})^{2}e^{- \left(\frac{{\Delta V_{\theta}}}{\Delta V_{\rm c}}\right)^{2}}\,.
\end{eqnarray}
$\Delta V_{\rm c} = 120\,{\rm km}\,s^{-1}$ represents an intermediate transition between the small 
galaxies with Gaussian HI profiles in which the velocity contributions add quadratically and giant 
galaxies with a `boxy' profile reproduced by the linear addition of the velocity terms. 
$\Delta V_{\rm t}\approx 20\,{\rm km}\,s^{-1}$ is the velocity width due to random motions in the disk~\citep{Rhee96,VS}. 

With this definition of $\theta$, zero corresponds to face-on and $\theta=\pi/2$ to edge-on. In 
cases where $\Delta V_\theta>>\Delta V_{\rm c}$, one can see that 
$\Delta V_\theta=\Delta V_{\rm t}+\Delta V_{\rm e}\sin\theta$. For $\theta=0$, one finds that 
$\Delta V_\theta=\Delta V_{\rm t}$, in other words the HI dispersion in the disk, whereas for 
$\theta=\pi/2$ we recover $\Delta V_\theta=\Delta V_{\rm t}+\Delta V_{\rm e}$ as expected.

In addition there is a broadening effect, $\Delta V_{\rm inst}$, of the HI profile due to the frequency 
resolution of the instrument, $R$. For a range of galaxy profiles, this broadening is found to be 
$\Delta V_{\rm inst}\approx 0.55R$~\citep{bot}. As befits a next generation ratio instrument the
ASKAP velocity width is extremely fine, $\Delta V_{\rm inst} \approx 4 \,\kms$, which is an insignificant source of 
error in the present discussion.

However, for completeness we add $\Delta V_{\rm inst}$ linearly to $\Delta V_{\theta}$, as 
argued by~\citet{hijass}, to give the effective observed linewidth,
\begin{eqnarray} \label{eq:effvel}
\Delta V_{\rm o}(\theta)= \Delta V_{\theta} + \Delta V_{\rm inst}\,.
\end{eqnarray}
The inclination angle $\theta$ can be solved for in Eqn~\ref{eq:TFq} by substituting the above effective observed linewidth 
(Eqn~\ref{eq:effvel}) and the intrinsic linewidth from Eqn~\ref{eq:velocity-mass relation}. We note that calculating the angle of inclination
from the matrix presented in Fig.~\ref{fig:HIPASS_W20_M} in this way finds less galaxies than by randomly assigning an angle of inclination, 
uniform in cosine, to the galaxies (which typically raises the completeness for WALLABY to $90\%$),
in this case the more detailed calculation is the more conservative estimate.

To determine the extent of an object on the sky, and hence ultimately the number of beams that resolve the structure, we make use of
an empirically derived relation between the HI mass of a galaxy and the HI diameter, $D_{\rm HI}$ (defined to be the region inside 
which the HI surface density is greater than $1M_{\odot}\,{\rm pc}^{-2}$). From~\citet{BR,VS} we have 
\begin{eqnarray}\label{eq:size mass relation} \frac{D_{\rm HI}}{\rm kpc}=\left( \frac{M_{\rm HI}}{10^{6.8}M_{\odot}} \right)^{0.55}\,.
\end{eqnarray}
The on-sky area of the galaxy can then be estimated~\citep{Meyer:08} using $\pi (D_{\rm HI}/2)^2 (B/A)^2$ where A and B are the major and minor axes 
respectively, the ratio of which (B/A) is equal to $\cos(\theta)$ which is calculated above. 
In practice we limit the smallest measurable angle of inclination for spirals to $\sqrt{0.12}$ in accordance with~\citet{Masters:03}. 
We compare the apparent area of the galaxy on the sky, scaling by the square of the angular diameter distance $d_{\rm A}(z)$, with the 
assumed Gaussian beam of ASKAP, $A_{\rm beam}$, given by 
\begin{equation}\label{eq:beamarea}
A_{\rm beam} = \pi \Omega_{\rm FWHM} / (4 \ln 2)\,.
\end{equation} 
where $\Omega_{\rm FWHM}$ was defined in Eq.~\ref{eq:beamdiam} previously.

As described at the start of Section~\ref{sec:method} we assume that the Signal-to-Noise of the galaxy is reduced when we are forced to 
recombine the multiple beams by which a galaxy is, potentially, resolved. 
The loss of signal is approximated by the square root of the number of beams needed to cover a given galaxy, given by the galaxy area, 
$A_{\rm gal}$, divided by the beam area, $A_{\rm beam}$. In practice the beam and galaxy widths are convolved when estimating this 
reduction, ensuring that even if the galaxy matches the beam size an additional factor of unity is added to this number of beams, giving a 
$\sqrt{2}$ reduction in signal in this case. Hence we reduce the peak flux $S^{\rm peak}$ in the detection Eq.~\ref{eq:peak limit} 
by this geometric factor $\sqrt{1+A_{\rm gal}/A_{\rm beam}}$.
For the current design of WALLABY, with a $2\, \rm km$ baseline, nearly $80\%$ of all galaxies are recovered after this reduction in peak flux.

\begin{figure}[!h]
  \begin{center}
    \epsfysize=2in
    \epsfxsize=4in
    \epsfig{figure=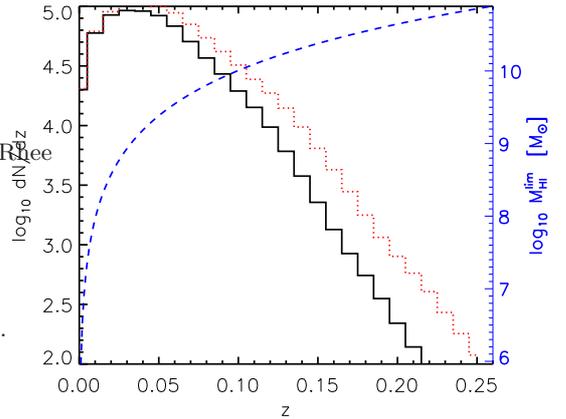, scale=0.45}
    \caption{In this figure we show the expected number counts of galaxies in redshift bins of
    		width $\Delta z = 0.01$, both with and without a loss of signal due to the resolving 
		out of galaxies by ASKAP (black solid and red dotted curves respectively). The effect
		is important, especially for systems at high redshift when the reduction in Signal-to-Noise is sufficient to push them
		below the detectability limit of ASKAP. The survey has values as described in 
		Table 1. WALLABY is approximately $80\%$ complete for a baseline
		of $2\rm \, km$. On the right axis, in blue, we plot the limiting HI mass as the dotted blue curve for a signal 
	      to noise detection of $5\sigma$ in one pointing in redshift bins of 
      	width $\Delta z = 0.01$ and a velocity width of $200\kms$.}
    \label{fig:ASKAP_mfn_dndz}
  \end{center}
\end{figure}

In Fig.~\ref{fig:ASKAP_mfn_dndz} we compare the predicted number counts as a function of redshift for the full sky WALLABY
survey both with (black, solid curve) and without (red, dot curve) the effects of resolving the galaxies. Clearly this effect
is particularly an issue for the faint distant sources which are both face-on and massive to be resolved out. 

\section{Galaxy survey}\label{sec:galsurvey}
In this section we combine our estimates of the detectability of galaxies from the previous section,
with the ASKAP strawman figures~\citep{Johnston:08} and the specifics of the WALLABY 
survey~\citep{WALLABY}, as summarised in Table 1. 
In Fig.~\ref{fig:ASKAP_mfn_dndz} the dashed blue curve indicates the expected neutral 
hydrogen mass limit as a function of redshift
in redshift bins of width $\Delta z = 0.01$ for a single pointing of ASKAP. The redshift depth of 
WALLABY is such that the survey ends when the mass limit approaches 
$\le 10^{11}\, \rm M_{\odot}$, which is the apparent maximal limit of HI systems. 

The expected number counts as a function of redshift on completion of the proposed survey
is shown in Fig.~\ref{fig:ASKAP_mfn_dndz} as the solid black curve
with the actual total number of detections and mean redshift of WALLABY given
in Table 1. 

\section{Cosmological Parameters}\label{sec:cosmo}
Using the predicted galaxy number counts for WALLABY we can estimate the errors on the galaxy 
power spectrum at the mean redshift of the survey $z=\langle z\rangle\approx 0.055$ and ultimately 
the expected constrains on cosmological parameters.
$P(k,z)$ is related to the power spectrum $P(k,0)$ by \begin{eqnarray}\label{eqn:pk}
P(k,z)=[b\,D(z)]^{2}P(k)\,,
 \end{eqnarray} 
where $b$ is the bias parameter and $D(z)$ is the growth factor computed\footnote{Using the excellent
publicly available {\it icosmo} package~\citep{icosmo}.} from
\begin{eqnarray} D(z)=\frac{5 \Omega_{m}}{2}E(z)\int^{\infty}_{z}\frac{(1+z^{\prime})dz^{\prime}}{[E(z^{\prime})]^3}\,,
\end{eqnarray} 
where $E(z)=H(z)/H_0$ and by construction $D(z=0)=1$. 

\begin{figure}[!b]
  \begin{center}
    \epsfysize=2in
    \epsfxsize=4in
    \epsfig{figure=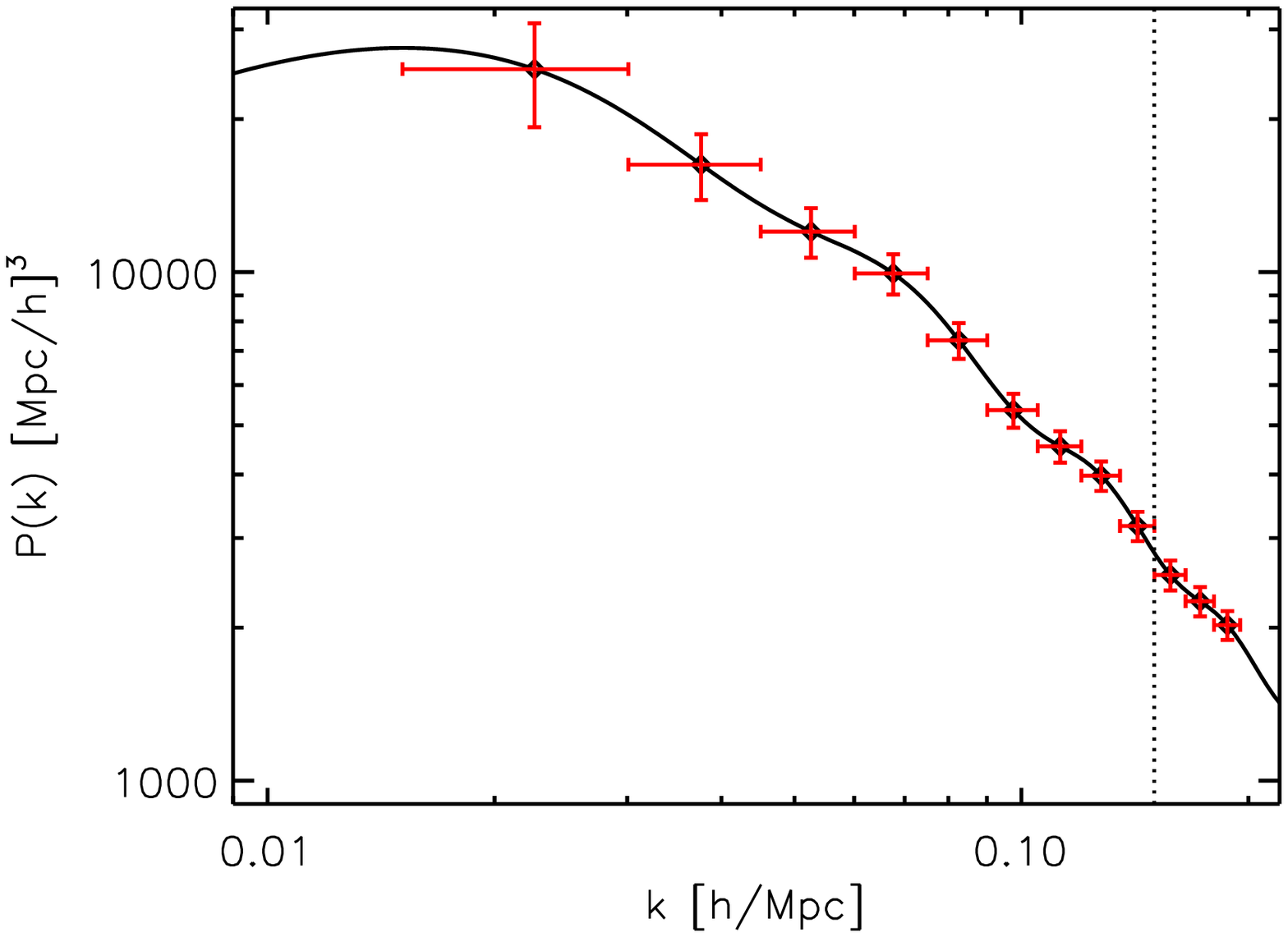, scale=0.45}    
    \epsfig{figure=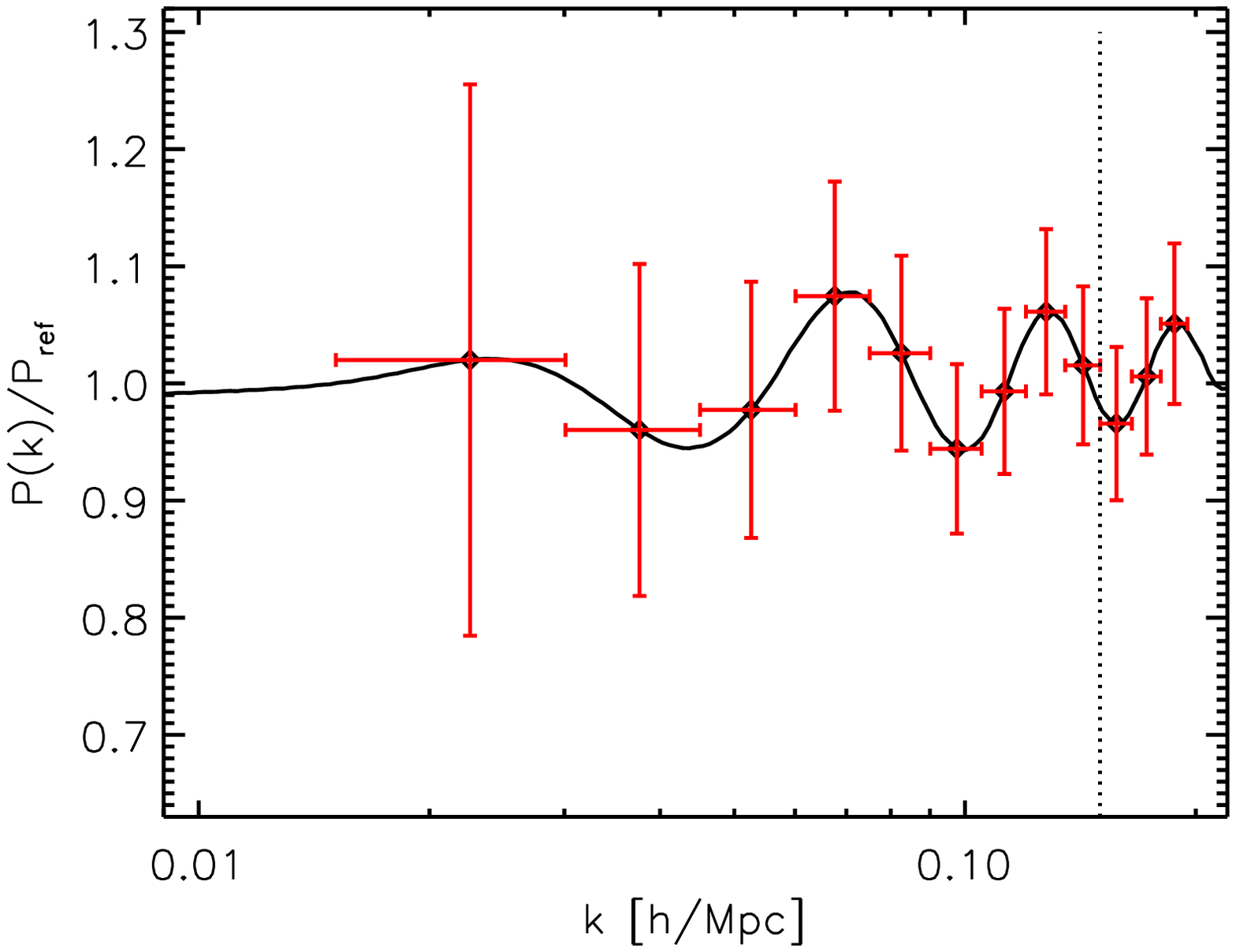, scale=0.45}
    \caption{The top panel shows the power spectrum (black curve) for the WMAP7 cosmology~\citep{WMAP7}, at the mean redshift
      of the WALLABY survey. The points are the expected errors on the measurement from the $3\pi$ sky survey with WALLABY. 
      The bottom panel is the same curve, normalised to a reference run with no baryons. The vertical dotted line indicates the smallest
      $k$-value ($k=0.15 \hmpc$) we consider for analysis.}
    \label{fig:ASKAP_P_k}
  \end{center}
\end{figure}

\begin{table}
\begin{center}
\begin{tabular}{cccc} \hline
Surveys & $w$ & $H_{0} $ \\ \hline
Planck &  $-1.02\pm 0.28$ & $73.3\pm 10.4$ \\ \hline  
Planck + WALLABY & $-0.98\pm 0.12$ & $71.2\pm 4.2$ \\ \hline 
Planck + 2dF &  $-1.00\pm 0.13$   & $72.0\pm 4.6$   \\ \hline 
\end{tabular}
\label{tab:planck}
\caption{Shown are the predicted cosmological parameter estimates
when projected {\em Planck} CMB data is used alone, then in combination with WALLABY
and also when used with an existing optical survey 2dF. Including the matter power spectrum measurements
results in a factor two improvement in the constraints on the Dark Energy equation of state parameter $w$ 
and the Hubble constant $H_0$.
Note that we performed a 6 parameter cosmological fit but that {\em Planck} has such small errors on most
parameters that a survey with less than a few $10^{6}$ sources is unlikely to 
improve the estimates, with the exception of $w$ and $H_0$, and for completeness we list the 
other variables which were unchanged when 2dF or WALLABY surveys were included;
[$\Omega_{\rm b}h^2 $,$\Omega_{\rm c}h^2 $,$n_{\rm s}$,$\log(10^{10}A_{\rm s})$,$\tau$] were
constrained to be [$0.0227\pm 0.0002$,$0.1099\pm 0.0015$,$0.964\pm 0.005$,$3.06\pm 0.01$,$0.092\pm 0.006$].}
\end{center}
\end{table}

Errors on the power spectrum are due to two factors: sample variance, i.e. the fact that not all $k$ 
modes are measured, and shot-noise, which is the effective noise on the measurement of an 
individual mode. The total error $\sigma_{\rm P}$ on the measurement of the power spectrum, 
$P(k,z)$, for a given $k$ with bin width $\Delta k$ can be expressed 
as~\citep{FKP,Tegmark1997,Blake:06}
\begin{eqnarray}\label{eq:Power error} \frac{\sigma_{P}}{P} =\frac{1}{\sqrt{m}}\left(1+\frac{1}{nP}\right)\,,\end{eqnarray}
where $P=P(k,z)$; $n=n(z)=\int_{M_{\rm lim}(z)}^{\infty}{dN\over dVdM}\,dM$ is the number density of galaxies which are detected 
(making $nP$ dimensionless)
and $m$ is the number of k-modes in a survey of total volume $V$, with $m= 2\pi k^2 \Delta k\, V/(2\pi)^3$.
The ability of a survey to probe cosmological parameters can be estimated, and compared, with the effective survey volume $V_{\rm eff}$ given by
\begin{eqnarray}
\label{eq:veff}
V_{\rm eff}(k)=\Delta \Omega \int_{0}^{\infty} \left(1+\frac{1}{nP}\right)^{-2} \frac{dV}{dzd\Omega}(z) dz\,.
\end{eqnarray}
As argued by~\citet{SeoEisenstein} there is no great gain in probing beyond $nP=3$ hence we limit the volume of our survey 
to a maximum redshift $z_{\rm max}$ when $n(z=z_{\rm max})P(k) = 3$; 
conservatively assuming a constant number density within this volume, hence the effective volume is now evaluated as
\begin{eqnarray}
\label{eq:veff_const}
V_{\rm eff}(k)=\Delta \Omega  \left(1+\frac{1}{n(z_{\rm max})P(k)}\right)^{-2} \int_{0}^{z_{\rm max}}\frac{dV}{dzd\Omega}(z) dz\,,
\end{eqnarray}
evaluated in Table 1 for two $k$-modes of interest ($k=0.065$ and $0.125$ $\mpch$), approximately corresponding
to the baryonic peaks in the power spectrum. Note that we have assumed a constant weighting function for the galaxies used in the power spectrum
measurement, a so-called `number-weighting' scheme. With our assumptions of constant number density within a given maximum redshift
the WALLABY survey is essentially a uniform survey  (i.e. with a window function which is effectively the identity matrix) and the covariance
matrix of the k-bands can be well represented by a diagonal covariance matrix with elements equal to $(4/3)^{2}(P^{2}/m)$ as argued in~\citet{Blake:06}.

In this work we create a power spectrum~\citep{Lewis:00} based on the latest WMAP7 Maximum Likelihood cosmology~\citep{WMAP7} for the standard
$\Lambda$CDM model with values $[\omb,\omm,\oml,h,w,n_{\rm s},\sigma_{8}]$ given by 
\noindent $[0.0451,0.271,0.729,0.703,-1,0.966,0.809]$. 
In the top panel of Fig.~\ref{fig:ASKAP_P_k} we demonstrate the power spectrum with the expected 
errors from WALLABY. In the bottom panel of this figure we have normalised the matter power 
spectrum by a reference no-oscillation power spectrum~\citep{EisensteinHu:98} to aid visualisation of the peaks.
It appears that with the errors from WALLABY it will be challenging to identify the `baryonic wiggles'. 
To quantify this we have performed a $\chisq$ test on the no-oscillation model with the datapoints in Fig.~\ref{fig:ASKAP_P_k} to determine if the 
datapoints have sufficiently small errorbars to rule out this wiggle-less model. The $\chisq / d.o.f$ value is $2.2 / 7$, this means that WALLABY
will only detect the BAO peaks with $\sqrt{2.2}\sigma$ significance. Although not conclusive this analysis does suggest that WALLABY 
will not significantly detect the Baryonic Acoustic Oscillations (BAO), 
therefore we do not attempt to use these oscillations as standard 
rulers\footnote{In~\citet{Beutler:11} they also predict that the BAO peak is marginally detectable in the correlation function using WALLABY, at $2.1 \pm{0.7}\sigma$
which is in agreement with our estimate}.

\begin{figure}[!h]
  \begin{center}
    \epsfysize=2in
    \epsfxsize=4in
    \epsfig{figure=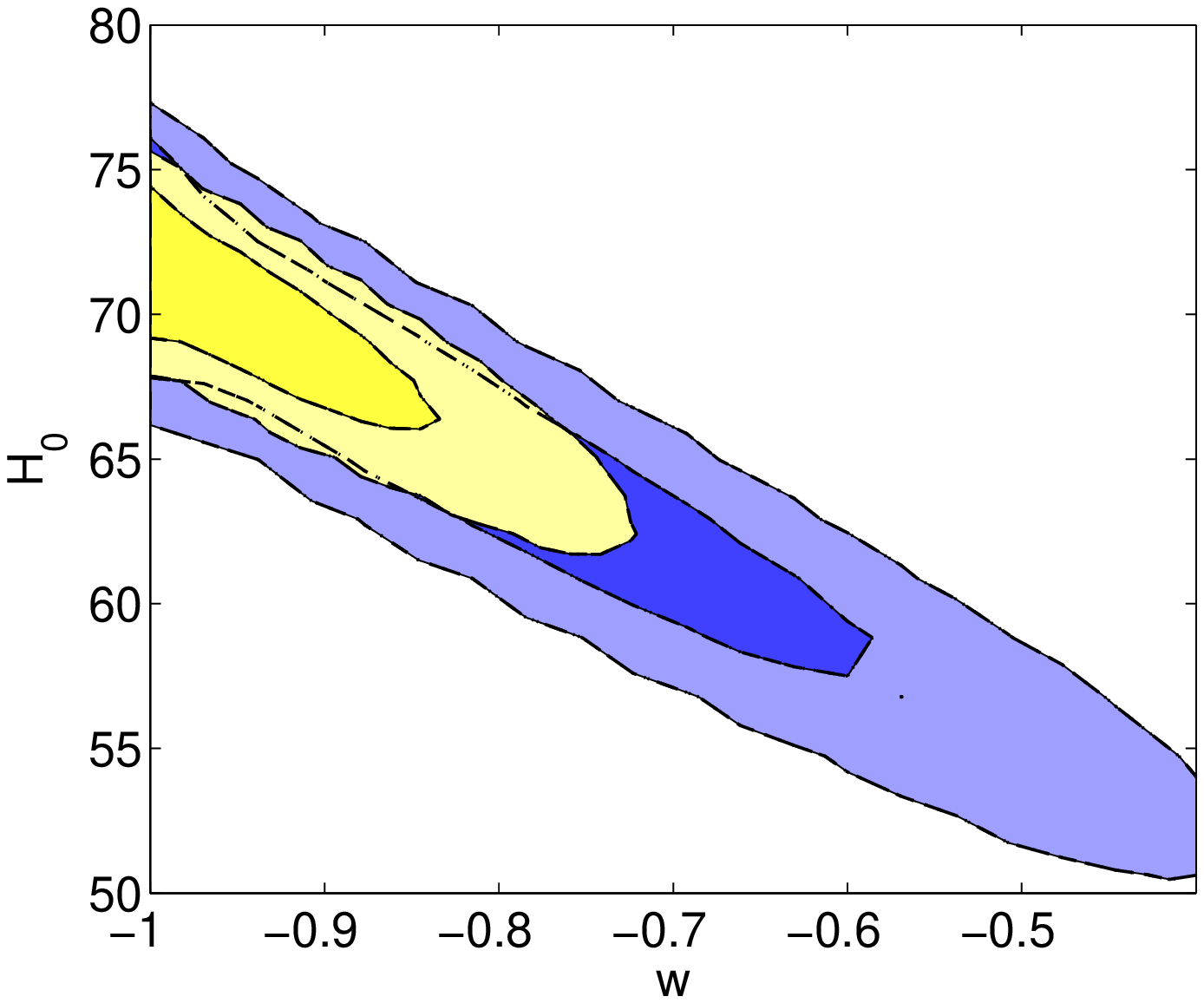, scale=0.5}
    \epsfig{figure=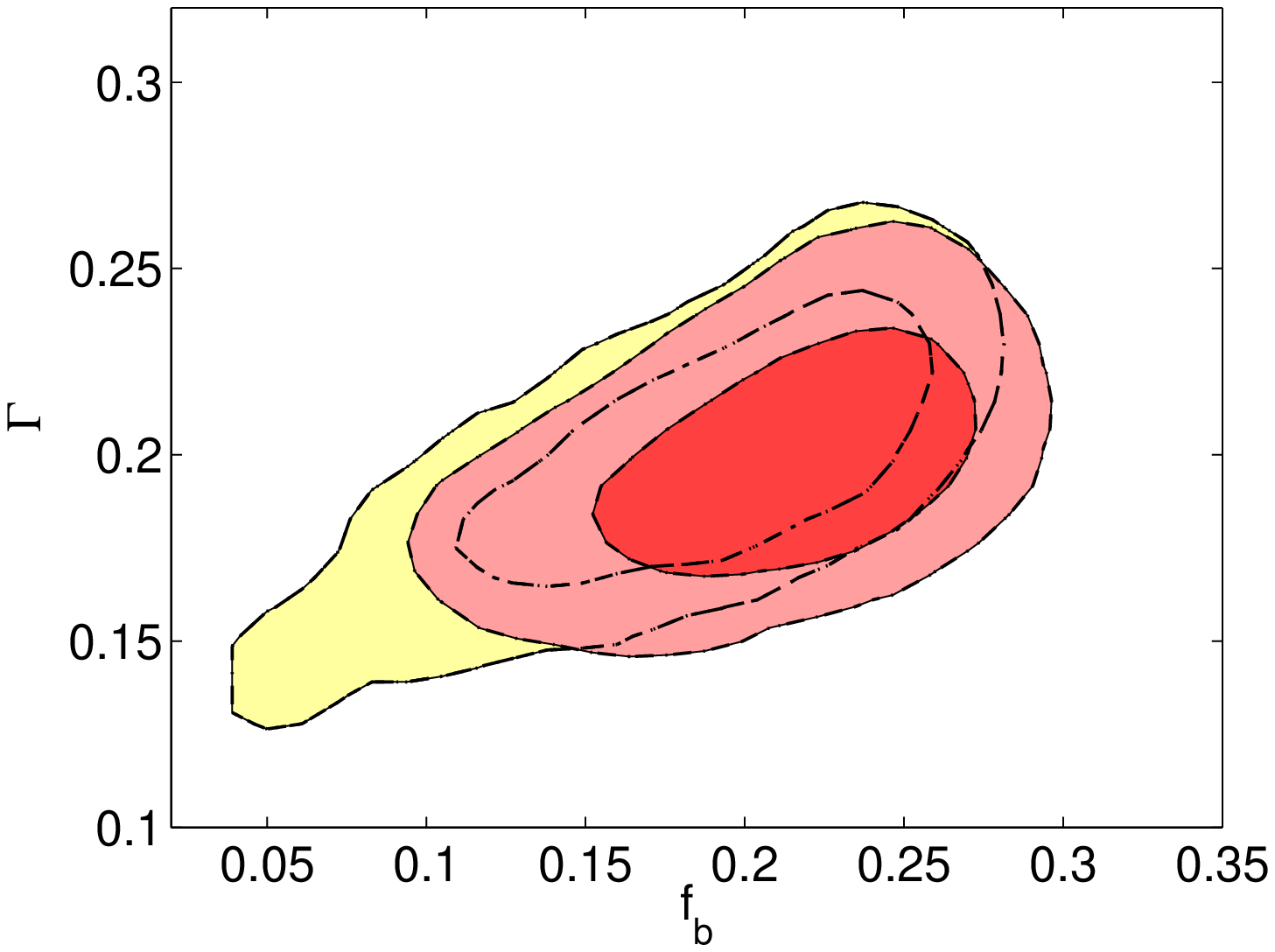, scale=0.5}
    \caption{Cosmological parameter error ellipses for degeneracies between the Hubble constant and the equation of state of Dark Energy
    in the top panel and the matter density in the Universe $\Gamma = \omm\,h$ versus the universal baryon fraction $\fb$ in the bottom panel.
    In the top panel we consider the case of a CMB measurement alone from {\em Planck} in blue and in yellow the improved constraint from the 
    measurement of the power spectrum by WALLABY (2dF gives a very similar result). 
    In the bottom panel we compare the results from the 2dF spectroscopic survey in
    red and the WALLABY estimates in yellow again. With these expected results ASKAP will likely be the first radio telescope 
    to successfully measure cosmological parameters from the matter power spectrum.}
    \label{fig:ASKAP_cosmo}
  \end{center}
\end{figure}

In accordance with our previous method of the analysis of FAST~\citep{Duffy:08a} we limit our investigation of the power spectrum to band-
powers over the range $0.005 < k/ \mpch < 0.15$. The maximum wavenumber chosen is a conservative cut on the power spectrum to ensure 
that we perform our analysis in the linear regime (as suggested in~\citealt{Cole:05}). It has been suggested (Beutler et al, {\it in prep}) that an 
HI survey such as WALLABY could safely trace even smaller scales ($\sim 0.2 \mpch$). This is because of the inherent low bias of HI detections 
which are tidally stripped in high-density regions, hence a blind HI survey will naturally avoid these regions, ensuring that the nonlinear
effects due to peculiar velocities of galaxies in these groups and clusters will be lessened, effectively extending the linear regime to smaller
scales.
For the Cosmic Microwave Background (CMB) data we have assumed that we have full polarisation information for {\em Planck} by 
considering the temperature T and E-type polarisation anisotropies for $l < 2400$ (including cross spectra), and assumed that they are 
statistically isotropic and Gaussian. 
The noise in the CMB data is also assumed to be isotropic and is based on a simplified model with 
$N_{TT} = N_{EE}/4 = 2 \times 10^{-4} \mu {\rm K}^2$, having a Gaussian beam of 7 arcminutes~\citep{Planck:06}. 
In the Markov-Chain Monte Carlo analysis we analytically
marginalise over the bias parameter $b$ in Eq.~\ref{eqn:pk} and assume that the priors around each cosmological parameter are flat, with 
a width safely outside that allowed by WMAP~\citep{WMAP7}.

The values shown in Table 2 are best-fit cosmological values~\citep{Lewis:02} for a variety of different parameters using 
expected {\em Planck} CMB data alone and then combined with WALLABY (or 2dF as a comparison).
By combining CMB data with the matter power spectrum measurement, which itself isn't a function of $w$, we break the degeneracy
between $w$ and $h$ that occurs when calculating the distance to the surface of last scattering from the CMB. Thus the 
main effect of ASKAP is to reduce the error on $h$ and $w$ by a factor two
on the value achieved with {\em Planck} alone, as demonstrated in the top panel of Fig.~\ref{fig:ASKAP_cosmo}. 
In the bottom panel of Fig.~\ref{fig:ASKAP_cosmo} we compare WALLABY with an existing optical based 
measurement of the matter power spectrum from 2dF, the two error ellipses are of similar size, graphically
illustrating the parameter estimates in Table 2 that WALLABY will be the first radio telescope to infer cosmological parameters
from the matter power spectrum.

\section{Conclusion}
As is clear from the galaxy survey estimates for WALLABY, ASKAP will likely be the first radio telescope to 
measure the matter power spectrum, constraining cosmological parameters to the level attained by the optical
survey 2dF in 2005, this will represent a coming of age for radio astronomy. 
To match, and ultimately surpass, current surveys such as WiggleZ will demand the full
version of the SKA. In creating a radio based galaxy survey the matter power spectrum will be analysed
using a different galaxy tracer together with different survey selection effects than the 2dF spectroscopic optical 
catalogue. Although not considered here there is the additional possibility of using the velocity field 
of the galaxies to gain additional cosmological constraints. As regards to a full local sample of 
$\approx 6 \times 10^{5}$ HI detected galaxies the science case is intriguing for the determination
of star formation in the local Universe. When coupled with deeper surveys 
on ASKAP, such as DINGO, that can determine the evolution in redshift out to 
$z=0.4$, the combined output will be a significant dataset for years to come.

\section*{Acknowledgements}
The matter power spectrum was created using {\em CAMB} and the cosmological 
parameter constraints were computed with {\em COSMOMC}, programs generously 
supplied by Anthony Lewis. The {\em ICOSMO} team should also be applauded for their
work in improving the ease with which cosmological parameters are calculated. 
We also thank Martin Zwaan and Martin Meyer for 
helpful science discussions as well as making available the HIPASS velocity-mass matrix.
ARD would like to make a special thanks to Florian Butler and David Parkinson for very helpful science
discussions. Finally, we would like to extend our sincerest thanks to Chris Blake 
for extremely valuable suggestions and queries which have greatly improved this article.

\end{document}